\documentclass[journal]{IEEEtran}
\IEEEoverridecommandlockouts
\usepackage{cite}
\usepackage{amsmath,amssymb,amsfonts}
\usepackage{algorithmic}
\usepackage{graphicx}
\usepackage{textcomp}
\usepackage{xcolor}
\usepackage{float}
\usepackage{subfigure}
\usepackage{upquote}

\def\BibTeX{{\rm B\kern-.05em{\sc i\kern-.025em b}\kern-.08em
    T\kern-.1667em\lower.7ex\hbox{E}\kern-.125emX}}
\begin{document}

\title{Trustworthy Image Semantic Communication with GenAI: Explainablity, Controllability, and Efficiency}
\author{Xijun~Wang, Dongshan~Ye, Chenyuan Feng, Howard~H.~Yang, Xiang Chen, and Tony Q. S. Quek 
\thanks{X. Wang, D. Ye, and X. Chen are with Sun Yat-sen University, China; C. Feng is with EURECOM, France; H.~H.~Yang is with Zhejiang University/University of Illinois Urbana-Champaign Institute, Zhejiang University, China; T.~Q.~S.~Quek is with Singapore University of Technology and Design, Singapore.} 
\thanks{This work has been submitted to the IEEE for possible publication. Copyright may be transferred without notice, after which this version may no longer be accessible.}
}

\maketitle

\begin{abstract}
Image semantic communication (ISC) has garnered significant attention for its potential to achieve high efficiency in visual content transmission. However, existing ISC systems based on joint source-channel coding face challenges in interpretability, operability, and compatibility. To address these limitations, we propose a novel trustworthy ISC framework. This approach leverages text extraction and segmentation mapping techniques to convert images into explainable semantics, while employing Generative Artificial Intelligence (GenAI) for multiple downstream inference tasks. We also introduce a  multi-rate ISC transmission protocol that dynamically adapts to both the received explainable semantic content and specific task requirements at the receiver. Simulation results demonstrate that our framework achieves explainable learning, decoupled training, and compatible transmission in various application scenarios. Finally, some intriguing research directions and application scenarios are identified. 

\end{abstract}


\section{Introduction} \label{intro}

Unlike conventional digital communication systems that prioritize bit-level correctness, semantic communication emphasizes the conveyance of meaning between transmitter and receiver. This approach extracts and transmits intricate features, commonly referred to as semantic-level information, directly from the source content. This ``transmit after understanding" philosophy significantly reduces redundant data transmission, positioning semantic communication as a promising mainstream technique for next-generation (6G) communication systems \cite{6Gniko2022}. As visual content, including photographs, infographics, and videos, proliferates at an unprecedented rate, the need for effective image transmission within semantic communication systems becomes increasingly important. However, this surge in visual data presents unique challenges, primarily due to the content richness and inherent ambiguity of images. Addressing these challenges is crucial for advancing Image Semantic Communication (ISC) technologies, which hold the potential to revolutionize user experience and dramatically improve overall efficiency.

A significant breakthrough in this field came with the introduction of Deep Joint Source-Channel Coding (JSCC) by Bourtsoulatze et al. \cite{bourtsoulatze2019deep}. This approach utilizes a unified neural network at the transmitter and receiver for both image encoding and semantic extraction, demonstrating superior performance in image reconstruction quality. Building upon this preliminary work, researchers have developed several innovative models to address various aspects of semantic communication. To enhance the image reconstruction quality, the Deep JSCC-f model was proposed in \cite{kurka2020deepjscc} by incorporating a channel feedback mechanism. A model with a feature scaling module that adapts to channel conditions was developed in \cite{xu2021wireless}. Additionally, hybrid automatic repeat request (HARQ) was exploited in \cite{Jiang2022HARQ} to reduce the semantic transmission error.




\begin{figure*}[h]
\centering
\includegraphics[width=0.8\linewidth]{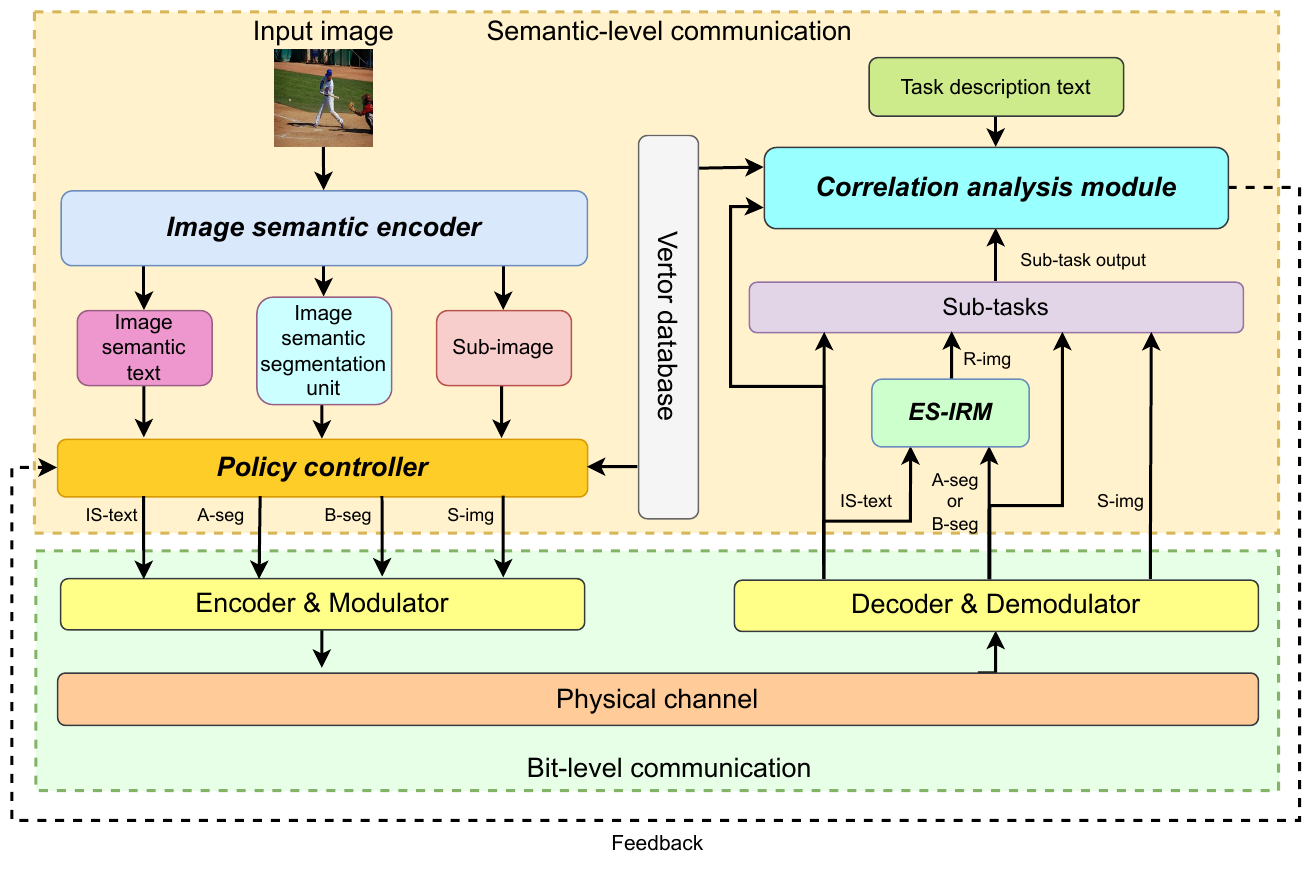}
\caption{An end-to-end trustworthy ISC framework based on system-compatible  explainable semantics. }
\label{fig-system}
\end{figure*}

Despite the promising advancements in deep JSCC-based schemes, several significant limitations persist, hindering their widespread adoption, trustworthiness, and effectiveness. These challenges include: i) lack of interpretability in semantic representations, as the extracted semantics are represented as opaque feature vectors, impeding algorithm optimization and human understanding; ii) operability issues in model training, where the joint training and simultaneous updating requirements for transmitter and receiver models complicate deployment, particularly for multi-downstream tasks; iii) compatibility problems in signal transmission, where the assumption of direct complex-valued signal transmission through channels often conflicts with current communication systems and hardware architectures; and iv) inefficiencies in multi-rate and multi-task scenarios, where current techniques fail to maximize transmission efficiency because of overlooking the semantics of source content.


In this article, we propose a novel trustworthy ISC framework designed to enhance explainability, controllability, and operability. Our framework features an image semantic encoder at the transmitter, which leverages pre-trained foundation models to convert images into explainable semantics through text extraction and segmentation mapping. At the receiver end, we employ cutting-edge Generative Artificial Intelligence (GenAI) to perform multiple downstream inference tasks, including image caption generation, image segmentation, and image reconstruction.
A key advantage of our framework is the elimination of the need for joint training or synchronous updates between transmitter and receiver models. The transmitter utilizes pre-trained models with robust generalization capabilities, reducing task-specific adjustments, while the receiver can independently update and train using pre-generated semantic data.
Furthermore, we introduce a semantic-level multi-rate ISC transmission protocol. This protocol is designed to maintain exceptionally high transmission efficiency while adapting to specific task requirements, ensuring optimal performance across various communication scenarios. 

The remainder of this paper provides a detailed exposition of our trustworthy ISC framework, presents encouraging results, explores potential applications, and concludes with future research directions.


\section{Trustworthy ISC Framework} \label{es-isc}

\begin{figure*}[ht]
	\centering
    \includegraphics[width=0.9\linewidth]{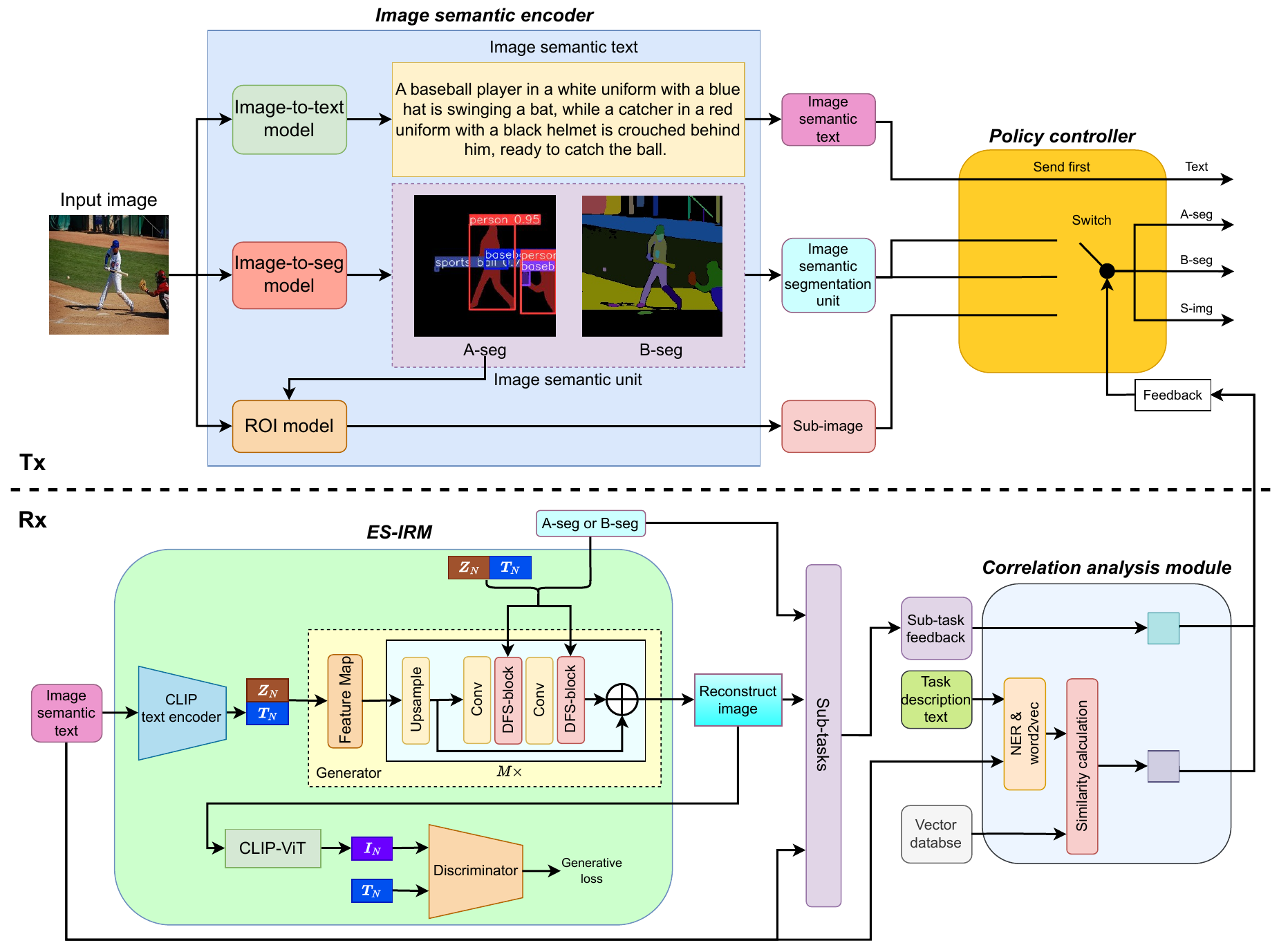}
	\caption{Details of key components in the trustworthy ISC framework, where A-seg and B-seg denote the segmentation maps based on semantic segmentation and Segment Anything, respectively, $T_N$ and $I_N$ denote the generated text and image feature vector, $Z_N$ denotes the additional noise for generative model training, NER stands for Named Entity Recognition, and DFS-block stands for Deep text-image-segmentation Fusion Block.}
    \label{fig-detail}
\end{figure*}

Our proposed unified end-to-end trustworthy ISC framework, illustrated in Fig. \ref{fig-system}, harnesses GenAI for multitask processing. The trustworthiness of this framework is underpinned by its explainability and controllability, which are ensured through two key aspects. Firstly, we employ explainable semantics as image semantic carriers, including image semantic text and image semantic segmentation maps. Secondly, the multi-rate transmission control is dynamically determined by both the received explainable semantic content and the specific task requirements at the receiver. This approach enables a flexible and transparent communication system adaptive to varying needs. The subsequent sections will delve into a detailed description of the main components.



\subsection{Explainability in Decoupled Transceiver Models}
Our framework incorporates an image semantic encoder at the transmitter to ensure compatibility with contemporary digital communication systems. This encoder, built upon a pre-trained foundation model, generates explainable semantics in discrete representation. At the receiver end, we implement an Explainable Semantics-based Image Reconstruction Module (ES-IRM). This module is designed to facilitate high-quality image reconstruction using the transmitted explainable semantics. Additionally, our framework also includes encoder and modulation module for transmitting explainable semantics through digital communication systems over physical channels. Similarly, semantics are recovered through decode and demodulation module at the receiver.


\subsubsection{Image Semantic Encoder} 
The image semantic encoder is designed to transform images into explainable semantics that facilitate downstream inference tasks. This encoder extracts two primary types of semantics: image semantic text (IS-text) and multi-level image semantic segmentation. As illustrated in Fig. \ref{fig-detail}, the latter encompasses three distinct levels of segmentation maps: A-seg, which provides a semantic segmentation map with object labels and instance boxes; B-seg, offering a comprehensive segmentation map with fine contours based on the segment anything model \cite{Kirillov2023ICCV}; and S-img, comprising sub-images of each object identified in A-seg, obtained through Region of Interest (ROI)-based masking operations on the original image. Importantly, all these semantic representations are generated in discrete data formats, ensuring seamless transmission over physical channels using modern digital communication systems. This multi-faceted approach to semantic extraction not only enhances the richness of the transmitted information but also maintains compatibility with existing communication systems.


\subsubsection{Explainable Semantic Image Reconstruction Module}
Our Explainable Semantic Image Reconstruction Module (ES-IRM) at the receiver leverages the Generative Adversarial contrastive Language-Image Pre-training (GALIP) model \cite{GALIP}, offering superior generation speed and computational efficiency compared to diffusion and autoregressive models. The ES-IRM's core architecture, illustrated in Fig. \ref{fig-detail}, harnesses the powerful image-text pairing capability of the Contrastive Language-Image Pre-training (CLIP) model \cite{CLIP}. It comprises a CLIP text encoder with its corresponding generator, and a CLIP image encoder (CLIP-ViT) paired with its discriminator. The process begins with the CLIP text encoder producing a text feature vector $T_N$, which, concatenated with a random noise vector $Z_N$, serves as input to the generator network. Concurrently, the CLIP-ViT generates an image feature vector $I_N$. The discriminator then utilizes both $I_N$ and $T_N$ to perform accurate similarity reasoning, validating the correspondence between the input image and text features. This mechanism guides the generator in producing text-consistent reconstructed images, ensuring a high-fidelity output that aligns with the transmitted semantic information. 


The generator network in our ES-IRM employs a sophisticated architecture comprising a feature map module and multiple Deep Fusion of text/image Segmentation (DFS) blocks, designed to enhance image reconstruction fidelity. Initially, the feature map module transforms the concatenated text feature vector and noise vector into an intermediate feature map. The DFS blocks, central to our approach, then take on the crucial task of image reconstruction based on this intermediate representation. These blocks fuse the text feature vector and received image segmentation maps into the reconstruction process, utilizing an affine layer \cite{tao2022df} and a Spatially-Adaptive Normalization (SPADE) layer \cite{SPADE}, respectively. This novel DFS-Block design significantly mitigates structure-level disparities between the generated and original images, ensuring a high degree of visual and semantic coherence. Consequently, our ES-IRM demonstrates remarkable capability in reconstructing high-quality images that  are visually similar and semantically close to the origin image.

\subsection{Controllability in Semantic-level Multi-rate Transmission}
Unlike conventional HARQ-based multi-rate semantic communication schemes that use uninterpretable feature vectors, our proposed approach achieves semantic-aware multi-rate transmission through three key components: a policy controller at the transmitter, a correlation analysis module at the receiver, and a shared vector database. This system facilitates adaptive and controllable data transmission, dynamically responding to both the semantic content of the communicated information and the specific requirements of the receiver's task.


\subsubsection{Correlation Analysis Module}
The receiver employs a correlation analysis module to evaluate the relevance of received semantics to the task at hand. This process begins with both the task description and image semantic texts being processed through a Named Entity Recognition (NER) module \cite{NER} to extract significant entity words. These extracted words are then converted into vector representations using a pretrained word embedding model. The module subsequently calculates the cosine similarity among the vectorized task descriptions, image semantic texts, and the vector database. Based on these similarity scores, the correlation analysis module determines whether additional sub-semantics should be requested from the transmitter.


\subsubsection{Policy Controller}
The policy controller at the transmitter strategically selects appropriate semantics for transmission based on feedback from the receiver regarding the type and status of downstream tasks. Initially, the image semantic text is sent due to its low communication cost compared to the entire image. The controller then iteratively selects which sub-semantics to transmit based on ongoing feedback until the task is reported as completed. For instance, only the image semantic text might be necessary for a question answering task, whereas an image reconstruction task might require the policy controller to selectively transmit semantic maps from A-seg or B-seg, or even sub-images. This flexibility enables efficient use of bandwidth while ensuring the receiver has sufficient information to complete its task.


\subsubsection{Vector Database}
The transmitter and receiver share a vector database containing word embeddings for all classes extracted by the image semantic encoder in A-seg. These embeddings, generated using models like Word2Vec \cite{word2vec}, enable efficient computation of similarity between downstream tasks and image semantic texts. The use of word embeddings is beneficial for accurately handling synonyms and semantic relationships through similarity calculations in the vector space.



\section{Application Scenarios } \label{experiments}

\subsection{Multi-task Scenario}
To evaluate the performance of the proposed framework, the Common Objects in Context (CoCo) dataset is used, which is a large-scale image dataset for object detection, segmentation, and image captioning released by Microsoft team\footnote{http://cocodataset.org/}. 
For generating image captions task, the Bilingual Evaluation Understudy (BLEU) score is used to primarily measure the similarity between two sentences. A higher BLEU score indicates closer approximation of the generated captions to the manually annotated ones. For image reconstruction task, three commonly-used performance metrics are employed: Peak Signal-to-Noise Ratio (PSNR) to indicate the pixel-level consistency of the reconstructed image, Learned Perceptual Image Patch Similarity (LPIPS) to measure the content consistency of the reconstructed image, and Fréchet Inception Distance (FID) to assess the diversity and distribution consistency between the reconstructed and original image sets. 

In conventional digital communication, the image is compressed firstly, then encoded and modulated, and finally sent to the physical channel. The receiver obtains the compressed content after demodulation and decoding, and then reconstructs the original image. This study employs a traditional digital scheme comprising JPEG compression,  Low-Density Parity-Check (LDPC) coding, and Quadrature Phase Shift Keying (QPSK) modulation. Two state-of-the-art algorithms, namley, SPADE \cite{SPADE} and GALIP \cite{GALIP}, are used as comparison schemes. GALIP utilizes textual descriptions for image generation, while SPADE focuses on image reconstruction using only segmentation maps. Our proposed method, ES-IRM, is evaluated in two distinct variants to assess its performance and versatility. The first variant, ES-IRM-A, integrates ES-IRM with image semantic text and exclusively utilizes A-seg maps for reconstruction. In contrast, ES-IRM-B follows a similar approach but employs B-seg maps instead of A-seg, allowing us to compare the effectiveness of different segmentation techniques within the ES-IRM framework.

\begin{figure*}[h]
\centering
\includegraphics[width=0.8\linewidth]{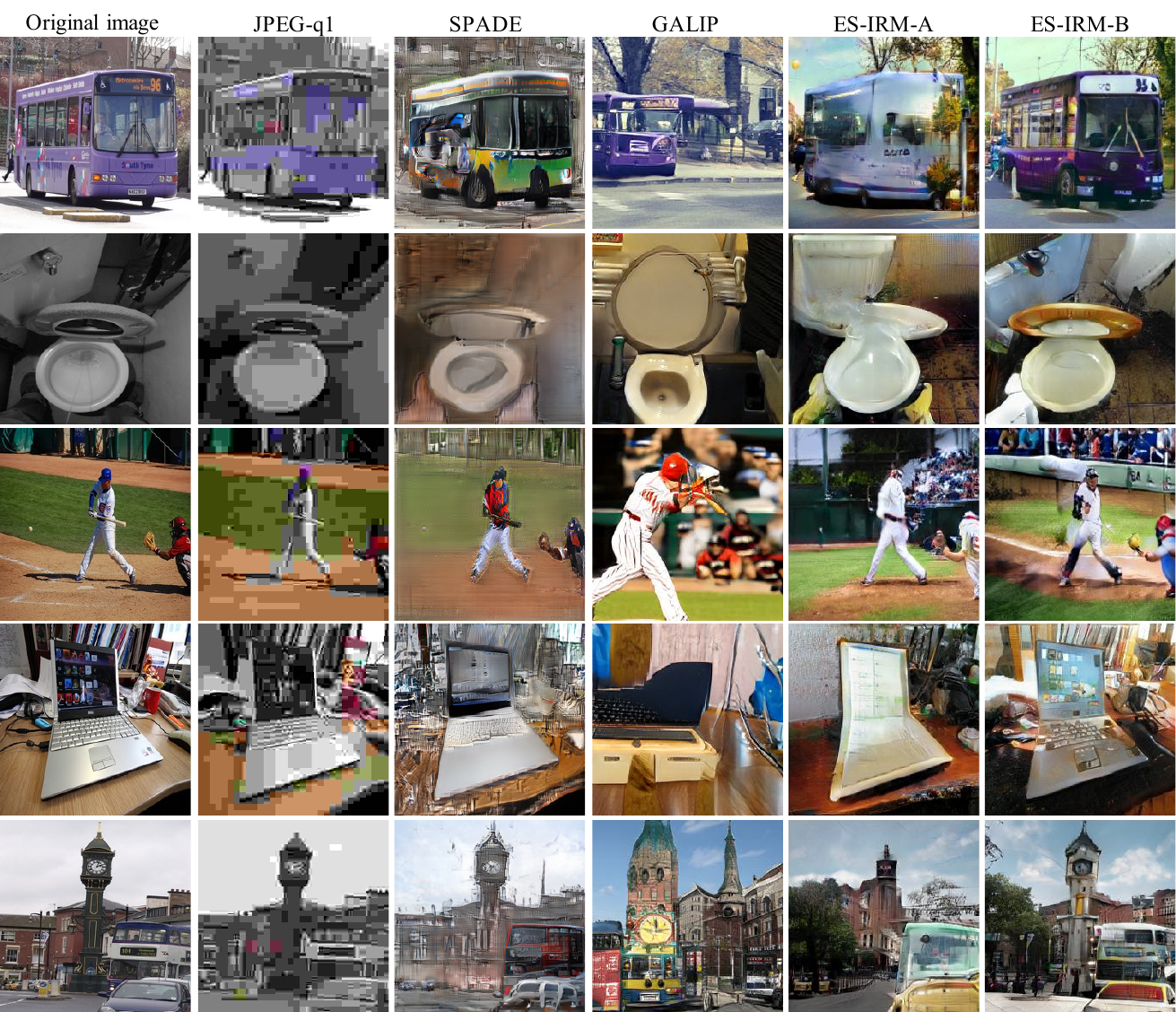}
\caption{Visualization of image reconstruction quality for various schemes. The first column contains the original images, and the remaining columns consist of the reconstructed images generated by JPEG-q1 (refer to JPEG compression algorithm at its lowest quality), SPADE \cite{SPADE} (refer to the semantic image synthesis algorithm with a  spatially-adaptive normalization), GALIP \cite{GALIP} (refer to the ingenious combination of CLIP and generative adversarial networks), ES-IRM-A (refer to an invariant algorithm that integrates ES-IRM with image semantic text and exclusively utilizes A-seg maps for reconstruction), and ES-IRM-B (refer to a similar approach but employs B-seg maps instead of A-seg), respectively. }
\label{fig-sample}
\end{figure*}

\subsubsection{Image Caption Generation Task}
In the context of image caption generation, our proposed semantic communication approach demonstrates significant advantages over conventional digital systems. While traditional methods compress images using JPEG at its lowest quality setting (JPEG-q1) to minimize data transmission, potentially causing severe visual distortion, our method transmits only the image semantic text. This allows the receiver to generate captions directly without the need for image reconstruction. Comparative analysis using BLEU scores on the test image set reveals a substantial performance gap: our semantic approach achieves a score of 33.613, while the conventional method only manages 15.898. This represents a 114.2\% improvement in caption quality, primarily attributed to the preservation of essential semantic information and the avoidance of visual distortions caused by aggressive JPEG compression in traditional systems. 


\subsubsection{Image Reconstruction Task}
Our proposed framework, particularly the ES-IRM-B variant, demonstrates superior performance in image reconstruction tasks, as evidenced by Table \ref{ES-IRM-vs}. ES-IRM-B achieves the best results in semantic-awareness metrics, specifically LPIPS and FID scores, where lower values indicate closer semantic similarity to the original image. Fig. \ref{fig-sample} visually confirms this superiority, showing ES-IRM-B's ability to closely replicate original image content, including specific details like a purple bus, a laptop, and buildings. In contrast, alternative methods exhibit significant limitations: JPEG-q1 suffers from severe color distortion and blocky artifacts due to extreme compression; SPADE struggles with complete object depiction (e.g., a purple bus), relying solely on segmentation maps; GALIP, using only semantic text, produces images drastically different from the originals (e.g., the reconstruction images of a baseball player swinging a bat and a laptop); and ES-IRM-A, while an improvement, lacks sufficient detail in some cases (e.g., a laptop). Overall, the proposed framework consistently generates more comprehensive and detailed reconstructions at comparable compression ratios, making it a promising solution for efficient image transmission and reconstruction.


\subsubsection{Transmission Efficiency}
Our experimental results highlight a stark contrast in communication requirements between various semantic representations. While image semantic text is typically limited to a mere 200 bytes, semantic segmentation maps and sub-images demand substantially more data. Specifically, A-seg requires an average of 1,952.09 bytes, achieving a 1/100 compression ratio relative to the original image, while B-seg needs 2,650.29 bytes on average, resulting in a 1.35/100 ratio. These ratios significantly outperform conventional Deep JSCC schemes, which typically achieve only 1/12 semantic compression. For comparison, JPEG at its lowest quality setting (JPEG-q1) requires an average of 2,094.93 bytes to reach a comparable 1/100 compression ratio. These findings underscore our approach's superior transmission efficiency while maintaining semantic integrity.


\begin{table}[h]
\centering
\caption{Image reconstruction performance}		
\label{ES-IRM-vs}
\renewcommand{\arraystretch}{1.5}
\begin{center}
\begin{tabular}{c|ccc}
\cline {1-4} 
Scheme & PSNR & LPIPS & FID\\
\hline \hline
JPEG-q1	    & 20.779     & 0.432    & 159.063\\
SPADE \cite{SPADE}	    & 11.507     & 0.534    & 63.601\\
GALIP\cite{GALIP}	    & 9.406     & 0.597	    & 39.846\\
ES-IRM-A (ours)	& \textbf{9.181}    & 0.552	    & 55.398\\
ES-IRM-B (ours)	& 10.375    & \textbf{0.403}	& \textbf{34.509}\\
\hline 
\end{tabular}
\end{center}
\end{table}

\subsection{Multi-rate Communication Scenario}

\begin{figure*}[h]
\centering
\includegraphics[width=0.8\linewidth]{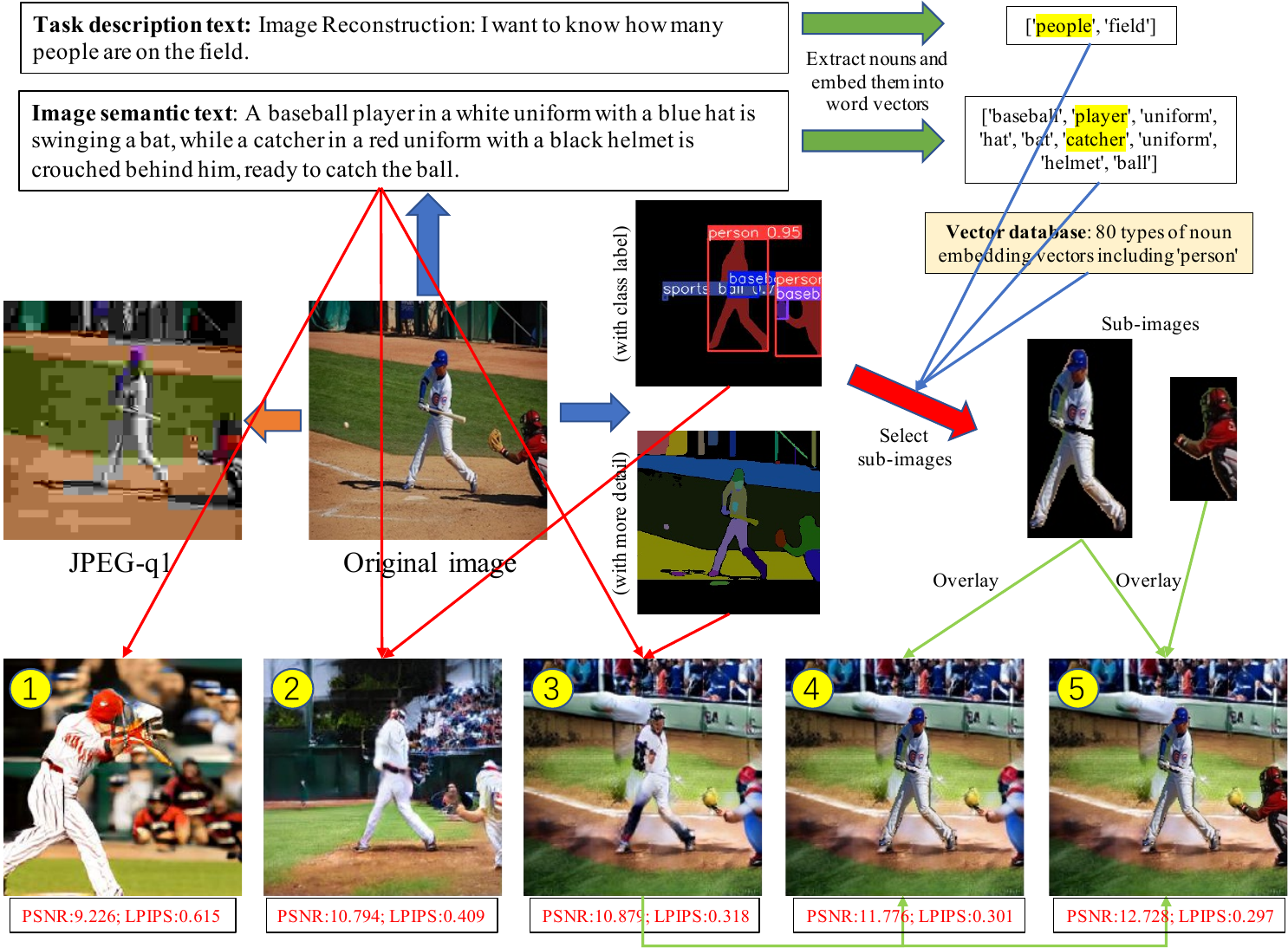}
\caption{Schematic diagram for single-user, single-task, and multi-rate communication.}
\label{fig-sample2}
\end{figure*}

\subsubsection{Single-user Single-task}
Fig. \ref{fig-sample2} demonstrates the efficacy of our proposed framework in a single-user, single-task multi-rate communication scenario. The receiver initially reconstructs an image using the received semantic text (the first image in Fig. \ref{fig-sample2}). If unsatisfactory, entity nouns from both task description and image semantic text are extracted and embedded as word vectors (e.g., [\textquotesingle people\textquotesingle, \textquotesingle field\textquotesingle] and [\textquotesingle baseball\textquotesingle, \textquotesingle player\textquotesingle, \textquotesingle uniform\textquotesingle, \textquotesingle hat\textquotesingle, \textquotesingle bat\textquotesingle, \textquotesingle catcher\textquotesingle, \textquotesingle uniform\textquotesingle, \textquotesingle helmet\textquotesingle, \textquotesingle ball\textquotesingle]). A correlation analysis between the vectorized task descriptions and image semantic text reveals semantic similarities, such as the word \textquotesingle people\textquotesingle in the task description corresponding to \textquotesingle player\textquotesingle\ and \textquotesingle catcher\textquotesingle\ in the image semantic text. Further analysis using the shared vector database identifies \textquotesingle person\textquotesingle\  as the related category among all A-seg categories.  Based on this analysis, the correlation analysis module requests additional sub-semantics related to the object label \textquotesingle person\textquotesingle\ from the transmitter to enhance image reconstruction quality. Subsequently, the transmitter progressively sends supplementary semantic information in the following order: A-seg maps focusing on the \textquotesingle person\textquotesingle\ category, detailed B-seg maps, and associated sub-images. This iterative process yields subsequent reconstructions (from the second to the fifth image) with improved PSNR and lower LPIPS scores. Our experimental results confirm that this multi-rate strategy enhances both the efficiency and quality of task completion progressively, showcasing its potential for adaptive, task-centric image transmission and reconstruction.

\subsubsection{Single-user Multi-task}
In our study of single-user, multi-task multi-rate communication, we simulate a scenario where a user must complete a diverse set of 480 tasks: 160 each for image captioning, segmentation, and reconstruction. These tasks are presented in random order to mimic real-world variability. We ensure task diversity by involving each category target in two distinct tasks. To create a comprehensive test dataset, we utilize Python's COCO API library functions to filter and select images from the COCO dataset. Specifically, we choose 50 images per category, resulting in a total of 4,000 test images. This selection method ensures that each category appears in approximately 400 images, providing a balanced representation across the dataset. We compare our proposed framework with three baseline schemes. The digital scheme combines the JPEG algorithm with traditional digital communication methods. As it lacks knowledge of specific downstream tasks, it uses a standard image compression rate (e.g., JPEG-q30) to ensure adequate quality for various potential tasks. The digital with knowledge scheme, while similar, leverages task awareness to optimize resource usage. It employs JPEG-q25 compression for image captioning while maintaining the digital scheme's approach for other tasks. Lastly, the ISC with knowledge scheme, grounded in explainable semantics and task awareness, selectively transmits specific types of sub-semantics tailored to each task.

Table \ref{SL-MR-ISC} illustrates the average data transmission requirements for each communication scheme. The digital scheme, employing JPEG compression at q30 for all tasks, requires 5,761.12 bytes per image. The digital with knowledge scheme reduces this for image caption tasks by using a lower compression rate. Our proposed framework and the ISC with knowledge scheme demonstrate superior efficiency for image caption tasks, requiring only 200 bytes of image semantic text. For image segmentation, leveraging the fact that each category appears in only 400 out of 4,000 test images, our correlation analysis module eliminates unnecessary data transmission for irrelevant images. Consequently, the average data required per image segmentation task is merely 395.21 bytes, with only 10\% of tasks needing both semantic text and A-seg maps. Similarly, an average of only 465.03 bytes is needed to complete an image reconstruction task. These results underscore the significant advantages of our proposed framework in multi-task scenarios, showcasing its ability to efficiently allocate communication resources based on specific task requirements and image content relevance.


\begin{table*}[h]
\centering
\caption{The average amount of communication data required to complete multiple tasks in multiple schemes (unit: bytes)}		
\label{SL-MR-ISC}
\renewcommand{\arraystretch}{1.5}
\begin{tabular}{c|ccc}
\cline {1-4}
&  Image caption task & Image segmentation task & Image reconstruction task \\
\hline
\hline
Digital scheme  & 5761.12     & 5761.12    & 5761.12\\
Digital  with knowledge	 scheme   & 4684.57     & 5761.12    & 5761.12\\
ISC with knowledge   & 200     & 1952.09	    & 2850.29\\
Semantic-level Multi-rate ISC	& \bf{200}    & \bf{395.21}	    & \bf{465.03}\\
\hline
\end{tabular}
\end{table*}


\section{Open Issues} \label{discussion}
Despite the promising advancements in image semantic communication, several challenges persist that require further research for widespread and effective deployment. This section outlines three critical areas that demand continued investigation and development.

\subsection{Device-adaptive Lightweight Deployment}
The challenge of ISC systems lies in device-adaptive lightweight deployment, particularly for Internet of Things (IoT) devices. These devices often have limited computational capacity, memory, and energy resources, making the implementation of ISC systems challenging \cite{light2023}. To address this, researchers are exploring attention mechanisms and edge computing as potential solutions. The core idea involves training semantic encoders and decoders on edge servers, then distributing the trained encoders to various terminal devices. This approach allows IoT devices to collect image data, extract semantic features, and transmit them to edge servers for further processing. Additionally, attention module is developed to assess the significance of each pixel and channel in image segmentation, thereby reducing the model's computational complexity. Lightweight AI design technologies, such as knowledge distillation and model pruning, are also being investigated to alleviate computational pressure on resource-constrained devices.

\subsection{Privacy-preserving Computation}
Privacy-preserving computation presents another significant challenge in ISC research. The primary goal is to ensure that sensitive information remains protected during data transmission and processing. Several strategies are being explored to address this issue, including homomorphic encryption, secure multi-party computation, differential privacy, and blockchain technology. For instance, blockchain can provide a decentralized storage and verification mechanism for sensitive image data, ensuring data integrity and immutability \cite{cao2023}. When combined with smart contracts, it can enable automated image processing tasks while maintaining transparency and security. Furthermore, trusted execution environments, such as Intel Software Guard Extensions, are being investigated to provide isolated execution environments that ensure the security of code and data during processing. These environments could allow key image processing operations to be executed securely, preventing unauthorized access to sensitive information.


\subsection{User-oriented Personalized Transmission}

Personalized service in ISC refers to tailoring the processing, analysis, and communication of image data to meet the specific needs and preferences of individual users or groups \cite{person2023}.  This user-oriented approach enhances user experience by providing customized responses and interactions based on the unique characteristics and context of the images being transmitted. Intuitively, creating detailed user profiles that capture preferences, interests, and historical interactions, could be incorporated to implement customization ISC services. Additionally, it is crucial to utilize context-aware computing to understand the situational context in which images are captured and used, allowing for more relevant semantic interpretations and responses. Establish feedback mechanisms that allow users to provide input on the personalized services they receive, which can be used to refine and improve the service. By implementing these strategies, personalized ISC services can greatly enhance user satisfaction and engagement, providing a more intuitive and responsive interaction with visual content, even for multimodal human-machine interactions in the near future.

\section{Conclusion} \label{conclusion}
In this article, we have investigated a trustworthy image semantic communication framework based on a decoupled transceiver model, utilizing explainable and system-compatible semantics. The framework employs image semantic text and image semantic segmentation maps as carriers for image semantics. Experimental results demonstrate that the proposed framework supports tasks such as image captioning, image semantic segmentation, and image reconstruction, outperforming traditional image compression and digital communication methods across all three tasks. Additionally, we have proposed a semantic-level multi-rate communication scheme. The experimental results show that it can achieve a 90\% reduction in the communication data volume required for semantic image communication. This scheme maximizes bandwidth utilization and enhances the overall efficacy of the communication system by enabling adaptive data transmission tailored to the specific requirements of the receiver.


\bibliographystyle{IEEEtran}
\bibliography{reference}

\end{document}